\documentclass[useAMS,usegraphicx,usenatbib]{mn2e}

\usepackage{graphicx}
\usepackage{times}
\usepackage{amssymb}
\usepackage{amsmath}
\usepackage{euscript}
\usepackage{longtable,lscape}

\def\210keV{{\rm\thinspace 2--10 keV}}
\usepackage{wasysym}
\title[Three AGN Close To The Effective Eddington Limit]{Three AGN Close To The Effective Eddington Limit For Dusty Gas}
\author[R.V. Vasudevan et al.]{R.V. Vasudevan$^1$\thanks{e-mail:ranjan@ast.cam.ac.uk}, A.C. Fabian$^2$, R. F. Mushotzky$^1$, M. Mel{\'e}ndez$^1$, L. M. Winter$^3$ \newauthor
and  M. L. Trippe$^1$ \\\footnotesize$^1$ Department of Astronomy, University of Maryland, College Park MD, 20742, USA \\\footnotesize$^2$ Institute of Astronomy, Madingley Road, Cambridge CB3 0HA, \\\footnotesize$^3$ Atmospheric and Environmental Research, 131 Hartwell Avenue, Lexington, MA, USA}

\begin{document}

\maketitle

\begin{abstract}
The Effective Eddington Limit for dusty gas surrounding AGN is lower than the canonical Eddington limit for hydrogen gas.  Previous results from the Swift/BAT 9-month catalogue suggested that in the overwhelming majority of local AGN, the dusty absorbing gas is below this Effective Eddington limit, implying that radiation pressure is insufficient to blow away the absorbing clouds.  We present an analysis of three objects from that sample which were found to be close to the Effective Eddington limit (NGC454, 2MASX J03565655-4041453 and XSS J05054-2348), using newly obtained XMM-Newton data.  We use the X-ray data to better constrain the absorbing column density, and supplement them with XMM optical monitor (OM) data, infrared \emph{Spitzer} and \emph{Herschel} data where available to construct a broad-band spectral energy distribution to estimate refined bolometric luminosities and Eddington ratios for these three objects.  The new XMM-Newton observations show all three objects moving away from the region expected for short-lived absorption in the $N_{\rm H}-\lambda_{\rm Edd}$ plane into the `long-lived absorption' region.  We find our conclusions robust to different methods for estimating the bolometric luminosity and Eddington ratio.  Interestingly, 2MASX J03565655-4041453 and XSS J05054-2348 now exhibit complex X-ray spectra, at variance with previous analyses of their Swift/XRT data.  We find evidence for absorption variability in NGC 454 and 2MASX J03565655-4041453, perhaps implying that although the radiation pressure from the central engine is insufficient to cause clearly detectable outflows, it may cause absorption variations over longer timescales.  However, more robust black hole mass estimates would improve the accuracy of the Eddington ratio estimates for these objects.

\end{abstract}


\section{Introduction}
\label{Intro}

The tight relation between the mass of the central black hole of a galaxy and the mass or velocity dispersion
of the surrounding stellar bulge points to a coupling or feedback between the black hole and the host galaxy.
The accretion power produced by the growth of the black hole exceeds the binding energy of the galaxy by
one to two orders of magnitude so some form of feedback is expected. What is unclear however is how it
works, and direct evidence is elusive and difficult to obtain. Radiation pressure from the Active Galactic
Nucleus (AGN) will have little effect on the stars of the surrounding bulge and only act on the gas. Catching
and recognising an object where the gas is being expelled has not been easy.
\cite{2009MNRAS.394L..89F} find clear evidence for the effect of radiation pressure from the central AGN on surrounding dusty
gas, using the most complete hard X-ray selected sample of AGN available at the time, the Swift/BAT 9-month catalogue (\citealt{2008ApJ...681..113T}). This catalogue consists of 97 low redshift AGN with measured column densities $N_{\rm H}$ and inferred black hole
masses.   They use the X-ray spectral fits of \cite{2009ApJ...690.1322W} (W09 hereafter) to measure the amount of gas, parameterised as the column density $N_{\rm H}$, and use X-ray luminosities in conjunction with a bolometric correction (to provide the full ionising continuum luminosity) and a black hole mass to parameterise the effect of radiation pressure, as the Eddington ratio $\lambda_{\rm Edd}$. The Eddington ratio $\lambda_{\rm Edd}$ is the ratio of the bolometric luminosity of an AGN ($L_{\rm bol}$) to the Eddington luminosity, or `Eddington Limit' for its black hole mass, $L_{\rm Edd} = 1.3 \times 10^{38}(M_{\rm BH}/M_{\odot})$.

\begin{figure}
    \includegraphics[width=8cm]{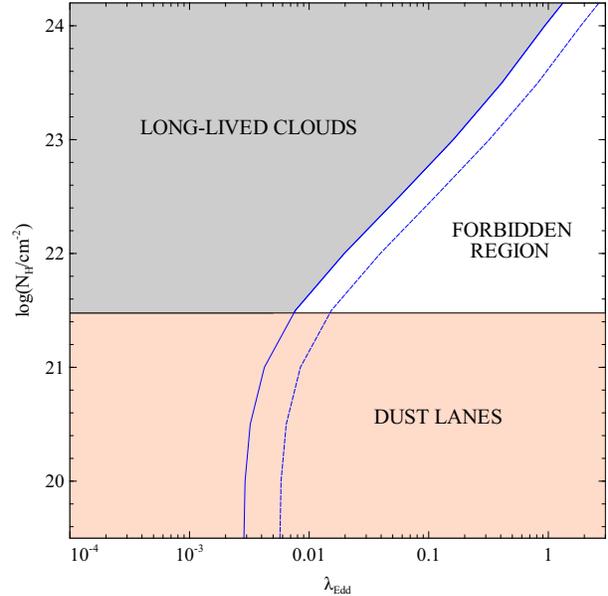}
    \caption{The Effective Eddington Limit paradigm: types of absorption expected if AGN are located in different areas of the $N_{\rm H}-\lambda_{\rm Edd}$ plane.  The Effective Eddington Limit, calculated using \textsc{cloudy} simulations, is shown by the blue solid curve (assuming standard interstellar medium grain properties), and differs from the canonical Eddington Limit ($\lambda_{\rm Edd} = 1$).  The dashed blue curve shows a factor of 2 increase in the effective Eddington limit due to the mass of intervening stars equal to the black hole mass itself (therefore providing a stronger gravitational counter-balance to radiation pressure for dusty gas clouds sufficiently far out from the nucleus).  AGN below this limit are expected to have long-lived absorption. AGN above this limit, in the `forbidden region' are expected to have short-lived absorption due to significant radiation pressure on the absorbing clouds, and should therefore not stay in this region for prolonged periods.  Column densities $\rm log(N_{\rm H})\lesssim 21.5$ may be due to dust lanes in the host galaxy. \label{effedd}}

\end{figure}

The canonical Eddington limit for Hydrogen and Helium relies only on electron scattering. When metals
and dust are mixed in cold gas, the cross section to radiation becomes much larger, due to the high absorption
cross section of dust in the UV, so the effective Eddington limit decreases. The deviation from the canonical
Eddington limit depends on the spectral energy distribution of the central AGN and the column density of
the absorbing clouds. Fig.~\ref{effedd} shows the variation of the effective Eddington limit as a function of $N_{\rm H}$ \citep{2008MNRAS.385L..43F}. The exact locus
of this effective Eddington limit also depends on details such as the dust-to-gas ratio of the surrounding
material, but at high column densities this has little effect. The radius of the clouds from the black hole
(BH) also alters their effective Eddington limit since the stars inward from the cloud will also exert inward
gravitational force, but even if the stars contribute as much mass as the central BH, this will only increase
the Eddington limit by a factor of two. This is represented by the blue dashed line in Fig.~\ref{effedd}.
In \cite{2009MNRAS.394L..89F}, results from the \emph{Swift}/BAT 9-month catalogue (with Eddington ratios
calculated using bolometric corrections from \citealt{2007MNRAS.381.1235V}) show a clear avoidance of the `forbidden'
region in the $N_{\rm H}-\lambda_{\rm Edd}$ plane, pointing to the pronounced influence of the AGN on its environment. Sources
well below the Eddington limit do not have sufficient radiation pressure to exert significant influence on their
host galaxies and so their absorption is stable. Absorbed sources at low column densities may appear above
their effective Eddington limit because of the presence of dust lanes in the galaxy (i.e. the mass interior to
the position of the clouds is much greater than the central black hole mass). The sources on the boundary,
near or above the effective Eddington limit, are of particular interest since they must be exhibiting an episode
of pronounced interaction between the AGN and the absorbing gas. The one source which lies well into the
forbidden region is MCG-05-23-16, which is known to have a complex and variable warm absorber (e.g. \citealt{2007PASJ...59S.301R}) and a possible variable, high-ionisation, high-velocity outflow \citep{2007ApJ...670..978B}, thus implying it is near the Eddington limit. For warm absorbers, the effect radiation pressure is
reduced. 
Detailed further study of other near-effective Eddington limited/super-effective Eddington sources such as these could be extremely
illuminating in understanding the role of AGN feedback in shaping the host galaxy.

There are three near-effective Eddington limited sources from the \cite{2009MNRAS.394L..89F} study which previously only had poorer-quality \emph{Swift}/XRT data - NGC 454, 2MASX J03565655-4041453 and XSS J05054-2348 - the XRT data were insufficient to constrain the detailed nature of the absorption in these sources.  We successfully proposed to obtain XMM-Newton observations for these three luminous absorbed AGN. These sources have complex optical spectra; for NGC 454, inspection of optical spectra from the REOSC Spectrograph at CASLEO 2.15 m Ritchey-Chretien telescope, San Juan, Argentina \citep{2000AJ....120..189D} shows a dual system with a Seyfert 2 (eastern source, NGC 454E) and a starburst galaxy (western source, NGC 454W); 6dF spectra for 2MASX J03565655-4041453 and XSS J05054-2348 reveal that these two objects can be classified as a starburst galaxy and a Seyfert 2 respectively (using the line ratio diagnostics of \citealt{1987ApJS...63..295V} and \citealt{2006MNRAS.372..961K}).

These three objects were previously observed for total periods of 20 ks (NGC 454, 345 counts), 12.6 ks (2MASX J03565655-4041453, 790 counts) and 3.2 ks (XSS J05054-2348, 389 counts) using XRT and the analyses of their spectra are presented in W09. For NGC 454, the XRT spectrum can be fit by a partial covering or double power law model giving a strong initial indication that complex absorption is at work, and the other two objects are fit by simple power-laws. However, given the low signal-to-noise ratio of the Swift/XRT data, such models may not be physically
appropriate to fully describe these systems and the acquisition of far superior quality XMM spectra allows us to fit more detailed models. The aim with this study is to constrain whether outflows are present and constrain the nature of the absorption (cold or warm/ionized). One of these XMM datasets (for NGC 454) has been analysed in detail by \cite{2012MNRAS.421.1803M}, and we build upon their conclusions in light of the Effective Eddington Limit paradigm here.  The XMM spectra allow more accurate determination of their X-ray luminosities and column densities, refining their position on the $N_{\rm H}-\lambda_{\rm Edd}$ plane. Determinations of source variability between different observation epochs will
be valuable in constraining how much these sources move around in this plane.
For these objects, the high column density makes it difficult to recover the bolometric luminosity from optical--to--X-ray SED integration, so we employ archival \emph{Spitzer} and \emph{Herschel} data where available to provide an alternative estimate, for calculation of Eddington ratios.  Additionally, we require good estimates of the black hole mass to determine the Eddington ratio.  We employ published black hole mass estimates using the black hole mass--K-band host galaxy bulge luminosity relation of \cite{2003ApJ...589L..21M} from the literature (although we note that the later study of \citealt{2007MNRAS.379..711G} provides an updated, more robust version of this relation for future determinations).  The approximate method outlined in W09 uses the archival 2MASS extended source magnitudes as an estimate of the total galaxy light (including the nucleus) and subtracts the published point source magnitudes for each source to remove the nuclear contribution.  However, \cite{2009MNRAS.399.1553V} (V09 hereafter) and \cite{2010MNRAS.402.1081V} (V10 hereafter) find that this over-estimates the bulge luminosity considerably, since the bulges of most of these galaxies are unresolved at the redshifts typical for the catalogue.  Therefore, they use the fainter point source magnitudes as the total `bulge plus nucleus' light, and use infrared SED templates to predict the level of nuclear contribution that needs to be removed in the K-band to get the bulge contribution alone.  We use the refined method of V09/V10 for our black hole mass estimates whenever possible.  It is not possible to obtain virial black hole mass estimates using optical broad lines for these sources (e.g., \citealt{2005MNRAS.356.1029B}) since the optical spectra for these sources indicate either Seyfert 2 or starburst types and do not display broad lines.

\section{Data Sources and Reduction}

The details of the XMM data used are presented in Table~\ref{table:datasources}. We reduce the raw data using the \emph{XMM-Science Analysis Software} (SAS) using the \textsc{epchain} and \textsc{emchain} tools as described in \cite{2009MNRAS.392.1124V}, including correcting for flaring by inspecting when the background lightcurve exhibited flaring episodes that were a significant fraction (up to a fifth) of the source flux and excluding those parts of the observation from further processing.  We extract the PN, MOS1 and MOS2 spectra using 36 arcsec source regions (corresponding to an encircled energy fraction of $\sim 85$ per cent for PN and MOS instruments) and multiple background regions near the source, using the standard procedures detailed in the XMM-SAS manual.  We group the final X-ray spectra with a minimum of 20 counts per bin, to allow $\chi^{2}$ statistics to be used in fitting the data in \textsc{xspec}.  We extract the optical--UV OM data using the \textsc{omichain} pipeline, checking for a good match between the optical source and the X-ray source position in each filter, and de-redden the OM fluxes for Galactic dust extinction using online dust extinction maps generated from the \cite{1998ApJ...500..525S} data\footnote{http://irsa.ipac.caltech.edu/applications/DUST/}, using the \cite{1989ApJ...345..245C} Galactic extinction curve.

\begin{table*}
{\begin{tabular}{l|l|l|l|l|l|l|l|l|l|l|l|l|l}

\hline
AGN & Redshift & RA & Dec & Observation & Observation & Observation       & Total counts & Usable $\%$ of obs.      \\
    &          &    &     & ID        & date & duration (ks)  & (PN and MOS) &(without flaring) \\
\hline
NGC 454 & 0.012158 & 18.605 & -55.397 & 0605090301 & 2009-11-05 & 29.9 & 4548 & 98 \\
2MASX J03565655-4041453 & 0.074782 & 59.235 & -40.696 & 0605090201 & 2009-09-12 & 39.5  & 9614 & 84\\
XSS J05054-2348 & 0.035041 & 76.441 & -23.854 & 0605090101 & 2009-08-06 & 30.9 & 25241 & 98 \\
\hline
\end{tabular}}
\caption{Table of \emph{XMM-Newton} observations used for each object. \label{table:datasources}}
\end{table*}

We supplement the optical, UV and X-ray data with infrared data from \emph{Spitzer} and \emph{Herschel} when available.  The \emph{Spitzer}/IRS spectra are available in the Cornell Atlas of Spitzer/IRS sources \citep{2011ApJS..196....8L}, and we find data for NGC 454 and XSS J05054-2348 in these archives.  We only find Herschel data for XSS J05054-2348; this object was mapped with the PACS \citep{2010A&A...518L...2P} and SPIRE \citep{2010A&A...518L...3G} instruments on board the \emph{Herschel} Space Observatory \citep{2010A&A...518L...1P}. Data were taken simultaneously with the PACS blue and red channels, centered on 70 and 160um. SPIRE observations were taken simultaneously at 250, 350, and 500um. The PACS and SPIRE data were reduced in standard way to level 1 using \textsc{HIPE} 9.0.0 \citep{2010ASPC..434..139O} and the final maps were made using the \textsc{Scanamorphos} software (Roussel 2012, submitted). A detailed description of the \emph{Herschel} data reduction process will be presented in Mel{\'e}ndez et al. (in prep).

\section{X-ray spectral fits (XMM PN/MOS and Swift/BAT)}
\label{xrayfits}

We fit the combined \textsc{pn}, \textsc{mos1} and \textsc{mos2} spectra along with the latest 58-month \emph{Swift/BAT} spectra, testing a suite of heuristic models using the \textsc{xspec} software to identify the best-fitting model, using the systematic method outlined in \cite{2013ApJ...763..111V} (Table 2).  In addition to attempting model fits using the model combinations in that paper (which include tests for Iron lines at 6.4~keV, soft excesses below 1~keV and edges due to highly ionised oxygen transitions at 0.73 and 0.87~keV), we also attempt other fits including a double power-law models (a direct component at high energies and scattered component at lower energies with the same photon index $\Gamma$, analogous to partially-covering absorption) and investigate whether low-energy residuals can be modelled as due to hot, diffuse gas using the \textsc{mekal} model.

We carefully order the different model fits based on their reduced chi-squared goodness-of-fit ($\chi ^{2}/\rm d.o.f$).  The best-fit model results for all three objects are presented in Table~\ref{fitresults}, where the errors on all quoted parameters are 90 per cent confidence intervals provided by the `error' command in \textsc{xspec}.

\begin{table*}
\begin{tabular}{llllllllllllllll}

\hline
AGN & $\Gamma$ (1) & $N_{\rm H}^{\rm int}$ (2) & $E_{\rm Fe K}$ (3) & $EW_{\rm Fe K}^{\rm PN}$ (4) & $L_{\rm 2-10keV}^{\rm unabs}$ (5) & Scattered & $\chi^{2}/\rm d.o.f.$ \\
    &          & ($\rm 10^{22} cm^{-2}$) &  (keV) &  (eV) & ($10^{43} \rm ergs^{-1}$) & fraction $(\%)$  (6) &  \\
\hline
NGC 454E & $1.87 \pm 0.06$ & $36.8^{+2.6}_{-2.3}$ & $6.34 \pm 0.05$ & $183^{+52}_{-102}$ & $0.1624 \pm 0.0003$ & 0.6 & 1.23 (384.2/313) \\
2MASX J03565655-4041453 & $1.63^{+0.13}_{-0.06}$ & $7.1^{+0.6}_{-0.4}$ & $6.43 \pm 0.06$ & $163_{-95}^{+93}$ & $3.54 \pm 0.02$ & 1.0 & 0.98 (681.0/694) \\
XSS J05054-2348 & $1.51^{+0.05}_{-0.04}$ & $9.9 \pm 0.3$ & $6.37 \pm 0.03$ & $115^{+50}_{-36}$ & $3.031 \pm 0.003$  & 0.8 & 1.06 (1286.0/1215) \\
\hline
\end{tabular}
\caption{Fit results for complex power-law best-fits to XMM-Newton and Swift/BAT data. (1): photon index. (2): absorbing column density. (3),(4): Energy/equivalent width of iron line as modelled by a redshifted gaussian (\textsc{zgauss}) component.  (6) scattered fraction: the ratio of normalisations of power-law components (low-energy scattered power law normalisation / high energy direct power law normalisation). This may represent an upper limit on the true scattered fraciton, since the soft emission may be contaminated by thermal, non-AGN emission. \label{fitresults}}
\end{table*}

\begin{figure}
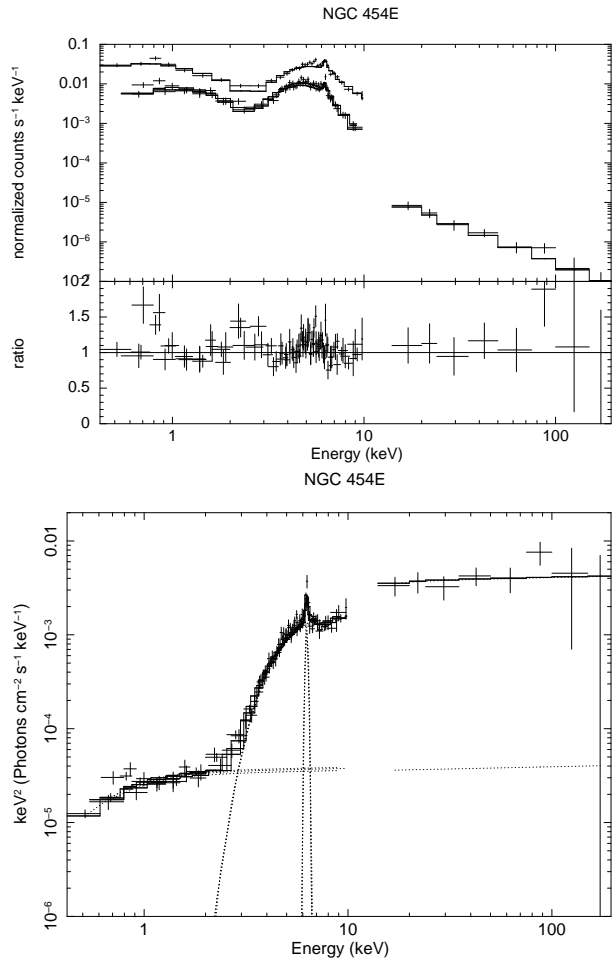

    \includegraphics[angle=270,width=8cm]{NGC454_bestfitmodel.ps}
    \includegraphics[angle=270,width=8cm]{NGC454_bestfitmodel_nuFnu.ps}
    \caption{NGC 454E: best fit model: an absorbed (Galactic and intrinsic absorption) double-power law with an Iron-line at 6.4~keV, reduced chi-squared: 1.23.  The upper panel shows the model fit to the data (in counts space) with a ratio plot, and the lower panel shows the unfolded spectrum in $\rm \nu \thinspace F_{\rm \nu}$ units.\label{ngc454spectrum}}

\end{figure}

\subsection{NGC 454}
\label{ngc454_xray}

In the XMM (PN and MOS) images, we see one prominent point source surrounded by a few faint sources, which can be ascribed to a number of different optical counterparts.  The work of \cite{1988A&A...191...29J} identifies optical components to NGC 454 identified as eastern (E), western (W), and three southern components (SE, S and SW).   The prominent X-ray source could potentially encompass emission from both the E and W components (the two merging galaxies), but analysis of the line ratios for the two sources presented in \cite{1988A&A...191...29J} would strongly suggest that the eastern component is responsible for the AGN emission, and is identified in \cite{1988A&A...191...29J} as a red elliptical galaxy in comparison to the western component, a blue, irregular (possibly disk) galaxy.  Analysis of Chandra images for NGC 454 (Mushotzky and Koss, private communication) also shows that the hard emission (above 2~keV) is centred on the eastern source, confirming its AGN nature.  We henceforth refer to this object as NGC 454E for clarity, to emphasise that this is the source responsible for the AGN emission of interest here.

The best-fit model is an absorbed double power-law with coupled photon indices (partial covering absorption) with an Iron line at 6.4 keV.  In \textsc{xspec}, the model prescription is \textsc{TBabs*zTBabs(powerlaw + zTBabs(powerlaw + zgauss))}, where the \textsc{Tbabs} component accounts for fixed Galactic absorption (determined using the \textsc{nh} utility from \textsc{fltools}) and the \textsc{ztbabs} components account for absorption intrinsic to the source; the \textsc{tbabs} model is adopted since it is the most up-to-date absorption model currently available in \textsc{xspec}.  The fit parameters are detailed in Table \ref{fitresults}.  We also attempt reflection and ionised absorption models, but do not find any significant improvement in the fit.  There are some prominent residuals at low energies ($\sim$1~keV) which could be identified as emission from hot, diffuse gas indicative of star formation contamination to the AGN spectrum from the western component.  The fit is marginally improved by inclusion of a \textsc{mekal} model to model these low-energy residuals (an improvement of $\Delta \chi^{2} \approx 7$). There are also unmodelled residuals near the Iron K-alpha line ($\sim 5-7$~keV) which have been attributed to an ionised absorber in \cite{2012MNRAS.421.1803M} (see also \S\ref{effeddparadigm}), but we do not attempt such a detailed fit here.

W09 find a column of $15.9\times10^{23} \rm cm^{-2}$ from XRT spectra, in comparison to $36.8^{+2.6}_{-2.3}\times 10^{22} \rm cm^{-2}$ found here using XMM.  The variation of a factor of $\sim 2$ suggest this is a `changing-look' AGN (e.g. see \citealt{2002ApJ...571..234R}), a scenario discussed extensively in \cite{2012MNRAS.421.1803M} by comparing the XMM-Newton dataset used here with \emph{Suzaku} and Swift/XRT data.   The intrinsic luminosity found here is very similar (within 13 per cent) of the value found in W09.  The \cite{2012MNRAS.421.1803M} analysis finds that the 2--10 keV intrinsic luminosity varies by a factor of $\sim$3--4 between XMM and \emph{Suzaku} observations along with a very pronounced variation in absorption (between $\sim 1 \times 10^{23} \thinspace \rm cm^{-2}$ and $\sim 2 \times 10^{24} \thinspace \rm cm^{-2}$, using their modelling approach outlined in Table~1 of their paper).

The BAT spectrum for this source is consistent with a power-law, and does not display any of the curvature expected from Compton reflection.  To check this, we fit the combined \emph{XMM} and BAT data with a simple \textsc{tbabs(pcfabs(pexrav))} model.  We ignore the Iron line region 5.5--7.5~keV, which is not modelled by \textsc{pexrav}.  This model includes Galactic absorption, intrinsic (partially covered) absorption and a reflection continuum parameterised by the photon index $\Gamma$, the reflection amplitude $R$ and the fold energy $E_{\rm fold}$ (in keV).  We freeze the other parameters in the \textsc{pexrav} model at their default values (abundances are solar, and the cosine of the inclination angle is fixed at 0.45), and check the validity of the fit by plotting contours of $R$ against $E_{\rm fold}$.  We allow the relative normalisations of the BAT and PN to vary with respect to each other, since the BAT spectra are not simultaneous with the XMM spectra.  This fit yields a reflection amplitude consistent with $R=0$ ($R<1.25$), a fold energy $E_{\rm fold}=56^{+140}_{-20}$~keV and a photon index of $\Gamma = 1.29^{+0.09}_{-0.04}$, much harder than the index of 1.87 obtained using the double power-law fit.   In these fits, there can be degeneracies when trying to independently constrain $R$, $\Gamma$ and $E_{\rm fold}$ all independently (see \citealt{2013ApJ...763..111V}); for example, the reflection and photon index can be correlated over a wide range of values if the high-energy cut-off is not well constrained.  Additionally, values of $\Gamma<1.5$ are not typical for AGN \citep{2011A&A...530A..42C} and are energetically difficult to obtain in standard Compton-scattering paradigms for the coronal emission \citep{1990ApJ...363L...1Z}.  We check the robustness of our reflection analysis by freezing the photon index at the value 1.87 in the \textsc{pexrav} fit and find that this results in a weakly constrained fold energy  (with only a lower limit of $E_{\rm fold}>106$~keV obtained from the error analysis) with a reflection fraction $13.9^{+8.8}_{-1.4}$, indicating that the uncertain relative normalisation of the BAT and XMM data makes it difficult to constrain reflection in this source.  The large value of reflection here would indicate a complex geometry; $R=1$ typically indicates a covering solid angle for the reflecting material of $2 \pi$ (representing, for example, an isotropic source above a disc), but values larger than this may be indicative of general relativistic effects such as light bending, or complex absorption, neither of which can be constrained with these data. \cite{2012MNRAS.421.1803M} also attempt a reflection fit to \emph{Suzaku} and \emph{Swift-BAT} data, but separate out the direct and reflected components rather than using a single \textsc{pexrav} model as done here.  They find a reflection fraction of $R \sim 1$ (i.e. a reflector covering $2 \pi$ steradians), consistent with what we find here, and notably show a very good agreement between the \emph{Suzaku} Hard X-ray Detector (HXD) data and the \emph{Swift-BAT} data (Fig.~4 of their paper).

\begin{figure}
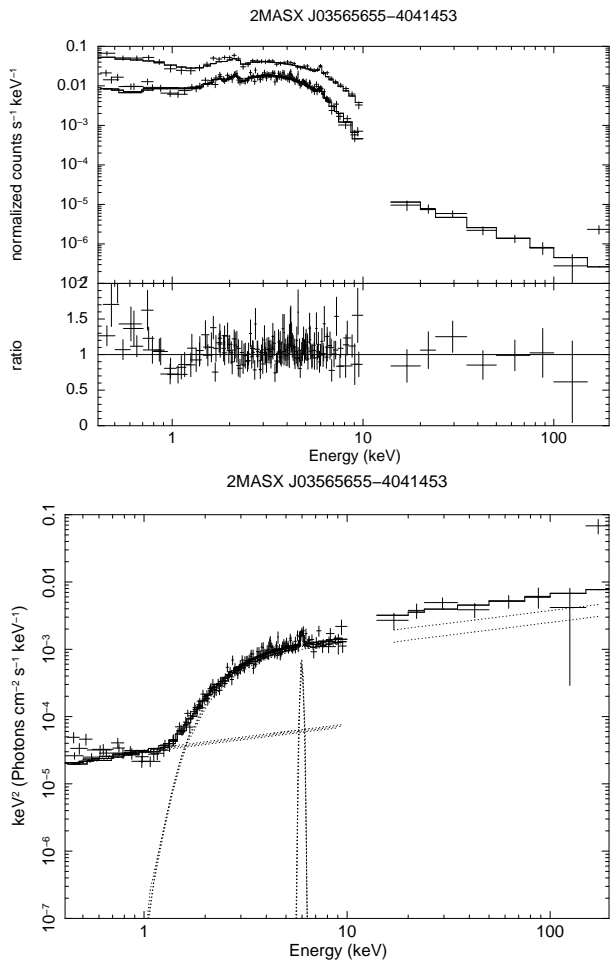

    \includegraphics[angle=270,width=8cm]{2MASXJ_bestfitmodel.ps}
    \includegraphics[angle=270,width=8cm]{2MASXJ_bestfitmodel_nuFnu.ps}
    \caption{2MASX J03565655-4041453: Best fit model:an absorbed (Galactic and intrinsic absorption) double-power law with an Iron-line at 6.4~keV, reduced chi-squared 0.98.  The upper panel shows the model fit to the data  with a ratio plot, and the lower panel shows the unfolded spectrum in $\rm \nu \thinspace F_{\rm \nu}$ units.\label{2masxj_xrayfig}}
\end{figure}

\subsection{2MASX J03565655-4041453}

The best-fitting model to the combined XMM and BAT data is again a double power-law with an Iron line at 6.4~keV (\textsc{TBabs*zTBabs(powerlaw + zTBabs(powerlaw + zgauss))}), as for NGC 454E (see Table \ref{fitresults} for details).  The absorbing column found here is $7.1^{+0.6}_{-0.4} \times 10^{22} \rm cm^{-2}$ which again represents an increase in a factor of $\sim 2$ over the absorbing column found in W09 using an XRT spectrum.  The intrinsic luminosity found here is about half that found in W09, suggesting significant intrinsic variability as well as a substantial variation in the absorbing column density.  Additionally, the XMM data reveal a complex spectrum (resembling partially-covered absorption) as opposed to the simple absorbed power-law shape found using the XRT data in W09.

We also perform a reflection analysis here using the \textsc{xspec} model combination \textsc{tbabs(zpcfabs(pexrav))}, as done for NGC 454E.  The shape of the BAT spectrum could be suggestive of reflection, and the best fit reflection parameters are well constrained: the reflection fraction is $R=3.4^{+2.0}_{-2.6}$, with a weakly constrained fold energy above 200~keV, and a photon index $\Gamma=1.84^{+0.06}_{-0.14}$, consistent (within 1$\sigma$ errors) with the value 1.63 obtained from the double power-law fit.  If we explicitly freeze the photon index at 1.63, we find a reflection fraction consistent with zero $R<1.32$ and a weakly constrained fold energy outside the BAT band ($E_{\rm fold}>182$~keV). Although the reflection fraction is consistent with zero (within errors) if we freeze the photon index, the better constrained photon index for this source may suggest that reflection is present on some level, potentially with a high reflection fraction.  A high effective reflection fraction could indicate light bending or extremely complex absorption in this source.

\begin{figure}
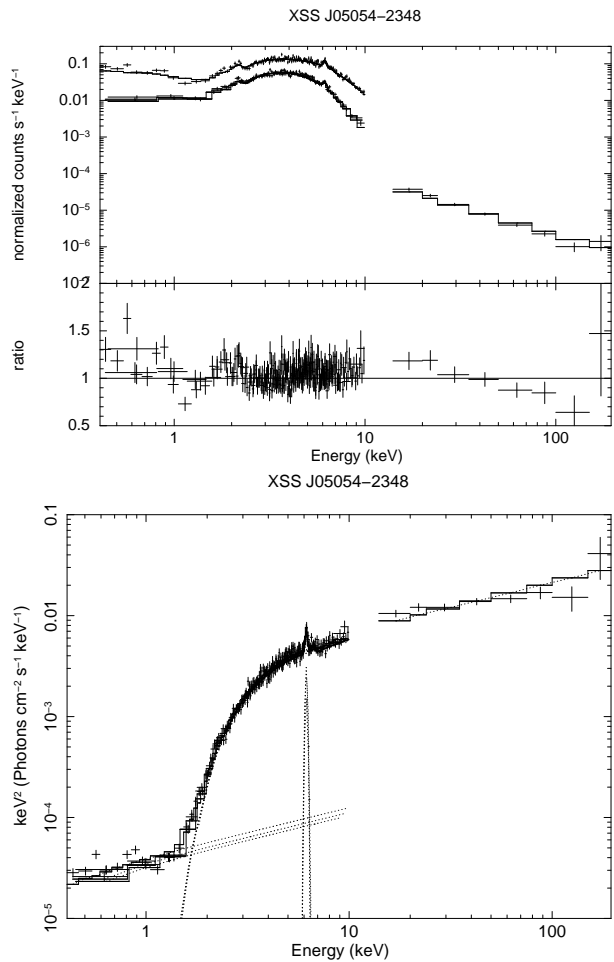

    \includegraphics[angle=270,width=8cm]{XSSJ_bestfitmodel.ps}
    \includegraphics[angle=270,width=8cm]{XSSJ_bestfitmodel_nuFnu.ps}
    \caption{XSS J05054-2348: Best fit model: an absorbed (Galactic and intrinsic absorption) double-power law with an Iron-line at 6.4~keV, reduced chi-squared: 1.06.  The upper panel shows the model fit to the data  with a ratio plot, and the lower panel shows the unfolded spectrum in $\rm \nu \thinspace F_{\rm \nu}$ units.\label{xssj_xrayfig}}
\end{figure}

\subsection{XSS J05054-2348}

A double power-law with an Iron line at 6.4~keV (\textsc{TBabs*zTBabs(powerlaw + zTBabs(powerlaw + zgauss))}) is again the best-fitting model for the combined XMM and BAT data (see Table \ref{fitresults}) for details.  We note the presence of some structure in the data below 2 keV, which can be well fit by a \textsc{mekal} model, improving the overall fit significantly ($\Delta \chi^{2} \approx 30$), but do not discuss this further here.  The absorbing column found here is $9.9 \pm 0.3 \times 10^{22} \rm cm^{-2}$, representing an increase in a factor of $\sim 1.5$ over the absorbing column found in W09 using an XRT spectrum.  The intrinsic luminosity found here is about 75 per cent of that found in W09, suggesting both moderate intrinsic luminosity and column density variation.  Here also, the XMM data reveal a complex spectrum (resembling partially-covered absorption) as opposed to the simple absorbed power-law shape found using the XRT data in W09.

A reflection analysis, using the same approach as for the other objects, yields the best fit reflection parameters $R=0$ ($R<0.05$), with a fold energy $E_{\rm fold}=44^{+11}_{-5}$, and a photon index $\Gamma=0.95^{+0.03}_{-0.03}$, much harder than the value 1.51 obtained from the double power-law fit.  This extreme photon index could indicate very complex absorption in this source.  Freezing the photon index at 1.51 obtained from the double power-law fit yields $R=11.9^{+0.9}_{-0.9}$ with $E_{\rm fold}=157^{+87}_{-52}$~keV, so there is considerable systematic uncertainty in the reflection parameters depending on the assumptions adopted.

\subsection{Comparison with archival XRT and Suzaku data}

\begin{figure}
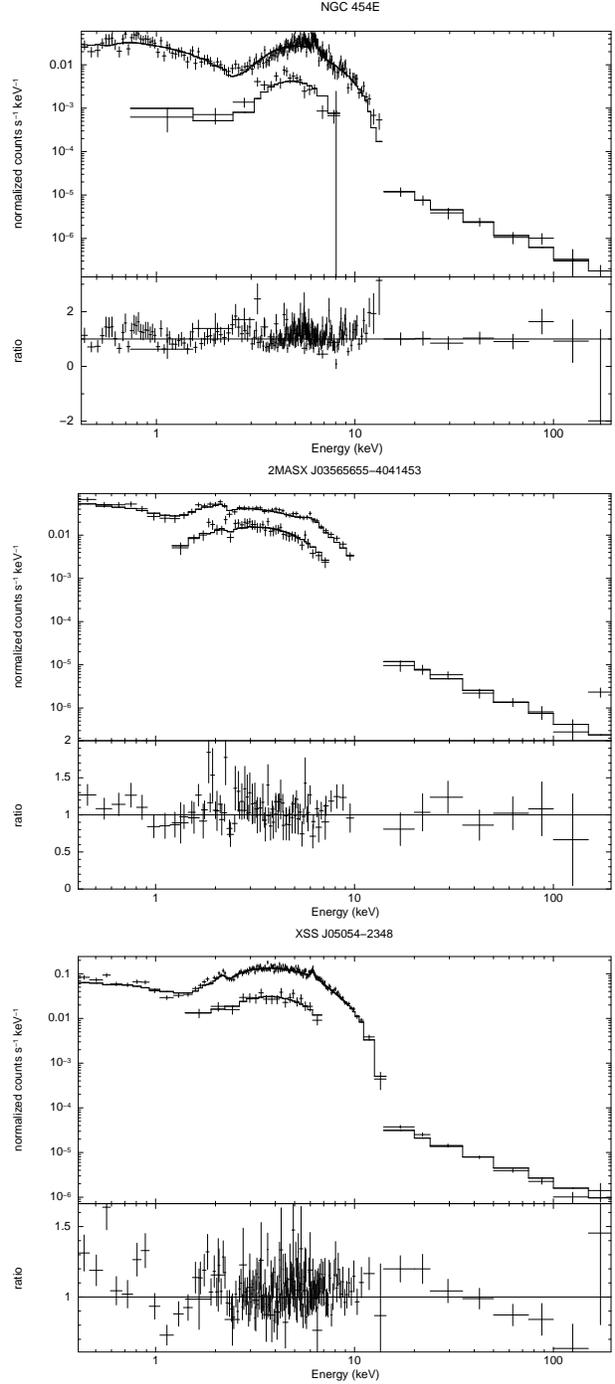

    \includegraphics[angle=270,width=8cm]{NGC454_XMMandXRTandBAT.ps}
    \includegraphics[angle=270,width=8cm]{2MASXJ_XRT_plus_XMM_plus_BAT.ps}
    \includegraphics[angle=270,width=8cm]{XSSJ_XRTplusXMMplusBAT.ps}
    \caption{Joint fits to both XMM (PN) and archival XRT data for the three objects.  The higher line below 10~keV in each plot is the XMM data (binned for 2MASX J03565655-4041453 due to high counts); the lower line below 10~keV is the XRT data (used in W09) and above 10~keV, we show the BAT data.\label{xrt_vs_xmm}}
\end{figure}

These three sources have all been studied in W09 using \emph{Swift}/XRT data.  We perform a brief comparison of our new XMM data with the XRT data used in W09, and fit the best fit model obtained for XMM+BAT data to XRT+BAT data.  For all three objects, the absorbed double power-law fit yields very similar parameters to the W09 fits to the XRT data alone (within errors), although the XRT data do not have sufficient counts to constrain the Iron line parameters.  NGC 454E has too few counts below 1~keV in the XRT spectrum to obtain a well-constrained photon index, and fitting the XRT data alone, W09 find $\Gamma=2.81^{+1.33}_{-0.88}$, whereas we find $\Gamma=1.91_{-0.47}^{+0.30}$ when including the BAT data.  We show a comparison between the new XMM datasets and the XRT datasets in Fig.~\ref{xrt_vs_xmm}, from fitting the double power-law model jointly to both datasets (we do this to purely show the degree of variation between the spectra when fit with a common model, and do not interpret the model fit results since it is not robust to draw physical conclusions from jointly fitting observations taken at different epochs).

\begin{figure}
    \includegraphics[angle=270,width=8cm]{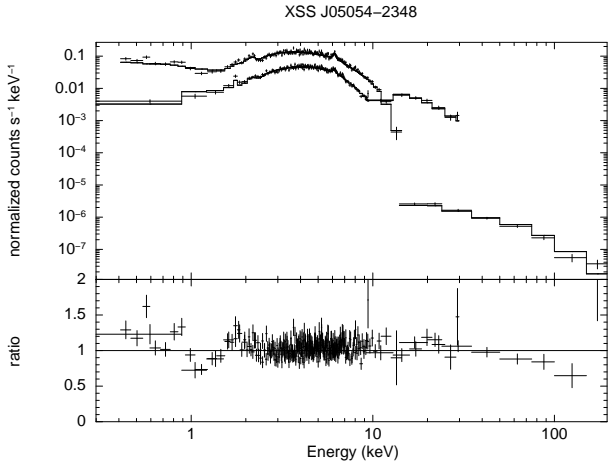}
    \caption{\emph{Suzaku}, XMM and BAT data for XSS J05054-2348, jointly fit with a double power-law with absorption and an Iron line.  The top line below 10~keV is the XMM data, the lower line below 10~keV is the \emph{Suzaku} XIS0 data, the higher line above 10~keV is the \emph{Suzaku} HXD data, and the lower line above 10~keV is the BAT data. \label{XSSJ_SuzakuandXMMandBAT}}
\end{figure}

An archival \emph{Suzaku} observation of XSS J05054-2348 has been studied in detail by \cite{2009ApJ...696.1657E}. They fit a complex, multi-absorber model to the combined \emph{Suzaku} XIS, PIN/HXD and \emph{Swift}-BAT data to obtain the absorption level in different components.  It is not straightforward to perform a direct comparison with their work due to the different modelling approach, but one of their absorbing components has a similar column density to the value found here (their value is $5.87 \pm 0.08 \times 10^{22} \rm cm^{-2}$, compared to $9.9 \pm 0.3 \times 10^{22} \rm cm^{-2}$ here, although the hard component in their study has a higher absorption of $29.3 \pm 6.2 \times 10^{22} \rm cm^{-2}$).  Regardless of the model fit, the shape of the \emph{Suzaku} spectrum is qualitatively very similar to that of the XMM spectrum taken three years later (see Fig.~\ref{XSSJ_SuzakuandXMMandBAT}); this is intriguing given that the \emph{Suzaku} light curve from 2006 shows very little variability compared to other sources in the \cite{2009ApJ...696.1657E} study.

\section{Eddington ratio estimates}

In order to refine the position of these three objects on the $N_{\rm H}-\lambda_{\rm Edd}$ plot, we also require more accurate determinations of the Eddington ratios.  Previously, in \cite{2009MNRAS.394L..89F}, Eddington ratios were calculated using a bolometric correction applied to the X-ray luminosity (to estimate the bolometric luminosity $L_{\rm bol}$), and dividing by the Eddington luminosity estimates calculated using the \cite{2009ApJ...690.1322W} black hole masses.   Here we refine this approach by constructing spectral energy distributions (SEDs) from infrared to X-rays, and consider the best approach for determining $L_{\rm bol}$ in these three AGN.  We make use of the XMM optical monitor (OM) data to construct the optical--to--Xray SED, but find that all three of our objects exhibit very reddened optical--UV spectra, probably due to significant dust in our line of sight or host-galaxy contamination.  This makes it difficult to recover the intrinsic ionising continuum responsible for exerting radiation pressure on the surrounding absorption.  If the significant X-ray absorption by neutral gas found in \S\ref{xrayfits} is coupled with dust as expected in AGN, this would also suggest that the red optical--UV SED shapes are likely due to significant dust reddening.  We therefore use two different techniques to estimate the bolometric luminosity: 1) we attempt to de-redden the optical--UV continuum for dust by including intrinsic reddening corrections in the optical--UV regime and fit an accretion disc and power-law model combination to to recover the full intrinsic luminosity from optical--to--X-ray wavelengths \citep{2009MNRAS.399.1553V}, and 2) we use the reprocessed infrared radiation (from mid-to-far infrared wavelengths, 8--1000 $\rm \mu m$) to estimate the portion of the intrinsic luminosity that is absorbed by circumnuclear dusty gas and re-emitted in the infrared and add it to the `remainder' of the intrinsic emission that appears at high energies above $\sim$10~keV (\citealt{2012MNRAS.425..623L}, \citealt{2010MNRAS.402.1081V}, \citealt{2007A&A...468..603P}).   We compare and contrast the results from both approaches, and determine whether the refined Eddington ratios for these objects still place them close to the Effective Eddington limit for dusty gas as found in \cite{2009MNRAS.394L..89F}.

\subsection{Bolometric luminosity}
\label{sec:bollum}

\begin{figure}
    \includegraphics[width=8cm]{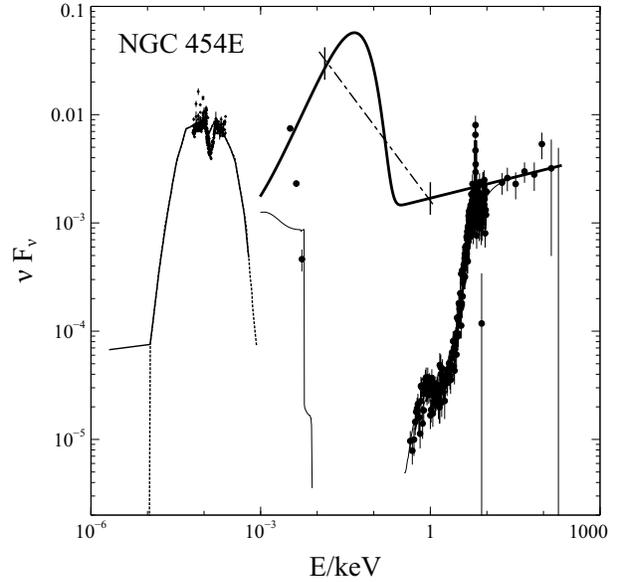}
    \caption{NGC 454E broad-band SED.  Large black filled points represent OM and PN data.  Above $10^{-3}$~keV, thin solid lines represent the model fits with absorption, heavy solid lines are absorption-corrected fits.  The long-dashed--short-dashed line linking 13.6~eV to 1~keV is the slope predicted using mid-IR high-ionisation line ratios (\citealt{2011ApJ...738....6M}, see \S\ref{highionIRlines} for details on this method).  Below $10^{-3}$~keV, small points represent \emph{Spitzer} data, the dotted line represents the nuclear IR emission component and the solid line represents the combined nuclear and host-galaxy emission (in this case dominated by the nuclear emission alone). \label{ngc454_sed}}

\end{figure}

\begin{figure}
    \includegraphics[width=8cm]{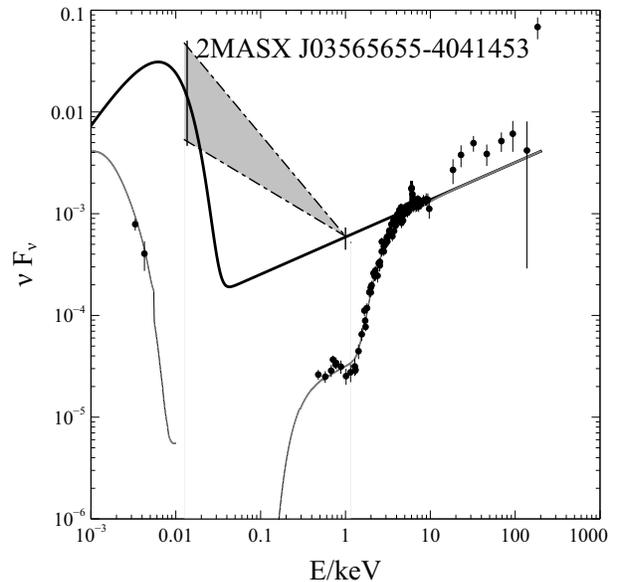}
    \caption{2MASX J03565655-4041453 broad-band SED.  Black filled points: OM and PN data.  Thin dotted lines represent the model fits with absorption, heavy solid lines are absorption-corrected fits.  The dot-dashed lines linking 13.6~eV to 1~keV show the range of 13.6~eV emission that can be obtained using the slopes predicted ($1.5<\alpha<2.0$ for $F_{\nu} \propto \nu^{-\alpha}$) using mid-IR high-ionisation line ratios (\protect\citealt{2011ApJ...738....6M}, see \S\ref{highionIRlines} for details on this method) found for the BAT catalogue.  For clarity, we only plot the model fits that assume the W09 mass, since the fit using the V10 mass is almost identical. \label{2masxj_sed}}
\end{figure}

\begin{figure}
    \includegraphics[width=8cm]{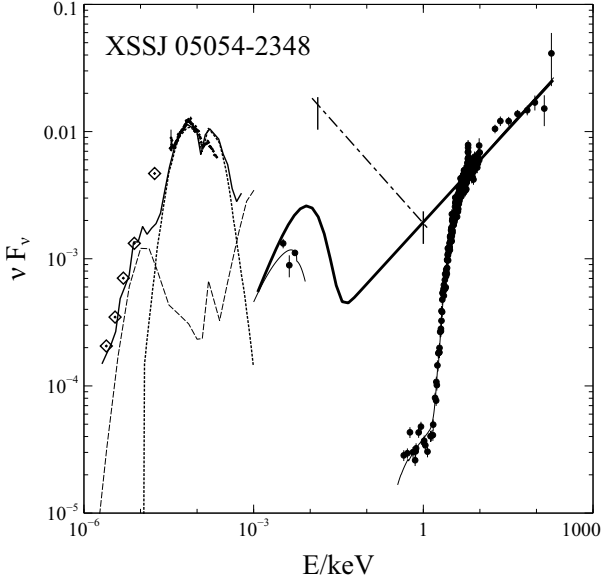}
    \caption{XSS J05054-2348 broad-band SED. Large black filled points represent OM and PN data.  Above $10^{-3}$~keV, thin solid lines represent the model fits with absorption, heavy solid lines are absorption-corrected fits.  The dot-dashed line linking 13.6~eV to 1~keV is the minimal slope $1.5$ predicted using mid-IR high-ionisation line ratios (\citealt{2011ApJ...738....6M}, see \S\ref{highionIRlines} for details on this method) found for the BAT catalogue.  Below $10^{-3}$~keV, small points represent \emph{Spitzer} data, diamonds represent \emph{Herschel} data, the dotted line represents the nuclear IR emission component, the dashed line represents the IR host-galaxy component and the solid line represents the combined nuclear and host-galaxy emission.  For clarity, we only plot the fits that assume the W09 mass, since the fit using the V10 mass is almost identical. \label{xssj_sed}}
\end{figure}

For using the optical--UV SED for determining $L_{\rm bol}$, we use the Optical Monitor (OM) data to construct SEDs in conjunction with the PN data for the three objects, and determine bolometric luminosities as detailed in \cite{2009MNRAS.399.1553V}, using black hole mass estimates to constrain the accretion disc model fit in the optical--UV.   We extract the magnitudes from the OM point sources as detailed in e.g. \cite{2006MNRAS.366..953B} and \cite{2009MNRAS.399.1553V} and construct the broad-band SED along with the PN data.   We also include the Swift/BAT data in the 14--195 keV energy range.  For the model fit, we use the best-fit found for the X-ray data alone as a starting point, and add in a \textsc{diskpn} component (a multicolour black-body accretion disc model) with intrinsic reddening incorporated using a \textsc{zdust} component to model the optical--UV accretion disc emission along with intrinsic reddening (assuming the `Small Magellanic Cloud' extinction curve).  We check all three sources for archival spectra with broad H$\beta$ and H$\alpha$ lines from which to estimate the intrinsic reddening E(B-V) \citep{1987ApJ...315...74W} but do not find any distinctive broad components.  We therefore produce an estimate of the intrinsic reddening by converting from the already determined X-ray $N_{\rm H}$ to a predicted dust extinction E(B-V) using a standard dust-to-gas ratio \citep{2001A&A...365...28M}.    For fitting the combined optical, UV and X-ray data, we identically follow the approach in \cite{2009MNRAS.399.1553V}.  The \textsc{diskpn} model has the following parameters: the inner radius of the disc $R_{\rm in}$, the maximal temperature in the disc $T_{\rm max}$ and the normalisation $N = M_{\rm BH}^{2} cos(\rm i) / D_{\rm L}^{2} \beta^{4}$ (for black hole mass $M_{\rm BH}$, inclination angle $i$, luminosity distance $D_{\rm L}$ and colour-to-effective temperature ratio $\beta$).  We assume $i=0$ and $\beta=1$ (as done in \citealt{2009MNRAS.399.1553V}) and $R_{\rm in}=6 \thinspace R_{\rm g}$ as is appropriate for efficient accretion.  We fit the optical--UV spectrum with the reddened accretion disc model combination to determine $T_{\rm max}$, which provides an estimate of the turnover in the unobserved far-UV.  The X-ray (XMM-PN) data were fit in conjunction with the optical--UV OM data, retaining the best-fitting X-ray model above 0.4~keV, giving a full model combination of \textsc{zdust(diskpn)+tbabs*ztbabs(powerlaw + ztbabs*(powerlaw + zgauss))}.  After a satisfactory fit was obtained, all absorptions (intrinsic-to-source and Galactic; both dust reddening and neutral gas) were finally set to zero before determining the `primary' (direct) estimate of the bolometric luminosity, defined as $L_{\rm bol}=L_{0.001-200 \rm keV}$.

For the second approach, we assemble data from \emph{Spitzer} where available, and supplement it with longer-wavelength \emph{Herschel} data where available.  We use the approach outlined in V10 and fit a nuclear and host galaxy SED template (using the templates from \citealt{2004MNRAS.355..973S}).  The nuclear SED templates are generated from a combination of mid-IR data and interpolation between data points using the dust emission model of \cite{1994MNRAS.268..235G}, and the host galaxy templates are generated using a combination of mid-to-far IR data (with the nuclear contribution subtracted), employing the \textsc{grasil} code\footnote{http://adlibitum.oat.ts.astro.it/silva/} \citep{1998ApJ...509..103S} for active and quiescent stellar populations to interpolate in the gaps not covered by the data. 

In this approach, different regimes of X-ray column density $N_{\rm H}$ are found to produce different IR spectral shapes for the nuclear emission, and different X-ray luminosities are found to produce different shapes for the host galaxy emission.  Using the X-ray determined $N_{\rm H}$ value, we select the appropriate nuclear IR template from the available ones provided in \cite{2004MNRAS.355..973S}, and we use the X-ray 2--10~keV luminosity to similarly select the appropriate host galaxy IR SED template.  We fit both templates jointly to the available near- and far-IR data. We determine the bolometric luminosity by integrating the nuclear IR component between 8 and 1000$\rm \mu m$, multiplying it by appropriate geometry and anisotropy corrections, and adding to the 0.5--200 keV \emph{absorbed} X-ray luminosity.  

The corrections for geometry and anisotropy are described fully in \citealt{2007A&A...468..603P} and V10 and are necessary to recover the full reprocessed emission.  The geometry correction corrects for the absorbing medium only partially covering the central accretion disc; it therefore only absorbs and re-emits part of the emission.  Based on the fraction of absorbed-to-unabsorbed AGN required by the \cite{2007A&A...463...79G} the covering fraction of the torus is estimated to be $f \approx 0.67$ and therefore the observed nuclear IR emission is scaled by $k_{\rm geom}=1/f$ to recover the primary emission (this value for the geometry scaling has been confirmed by the more recent analysis of \citealt{2010A&A...517A..11P}).  

The anisotropy correction takes into account the fact that the nuclear IR spectra of sources with different $N_{\rm H}$ differ significantly in the 1--30$\rm \mu m$ band, with higher $N_{\rm H}$ AGN exhibiting a notable deficit in this region due to anisotropic emission from the torus (see Fig.~1 of \citealt{2004MNRAS.355..973S}).  Under the unified model, lower $N_{\rm H}$ sources are more `face-on' to the observer, so to recover the true total emission from higher $N_{\rm H}$, more inclined sources, we apply a correction to recover the `missed' flux that escapes from our line-of-sight.  We adopt the Seyfert 1 SED template as the true spectral shape of a face-on AGN and normalise all other $N_{\rm H}$-group SED templates to have the same flux as the Seyfert 1 SED in the 30--100 $\rm \mu m$ band (where anisotropy differences are negligible for different $N_{\rm H}$ groups).   The correction factor for each $N_{\rm H}$ group is then computed by dividing the integrated flux of a Seyfert 1 SED template over 1--30 $\rm \mu m$ (where anisotropy differences are most pronounced) by the observed flux in the same band for each of the $N_{\rm H}$ groups (as done by \citealt{2007A&A...468..603P}). The correction factor increases with $N_{\rm H}$ and we adopt the same corrections as used in V10, i.e. $k_{\rm anis}=1.3$ for $22 < \rm log \thinspace N_{\rm H} \thinspace < 24$ sources and $k_{\rm anis}=3.5$ for $\rm log \thinspace N_{\rm H} \thinspace > 24$ sources.  We finally add the corrected IR nuclear emission ($L_{\rm 8-1000 \rm \mu m}^{\rm (obs)} \times k_{\rm geom} \times k_{\rm anis}$) to the emission that emerges at hard X-ray wavelengths (i.e., the `remainder' of the intrinsic emission that gets through the absorption, up to 200~keV) to determine the total $L_{\rm bol}$.   

We perform a Monte-Carlo analysis to determine errors at each stage, but find that systematic effects dominate and that the choice of method to determine $L_{\rm bol}$ prouces a much more pronounced variation on estimates than the formal errors would indicate.  We therefore quote ranges for $L_{\rm bol}$ and derived parameters, rather than providing formal errors.  All key results from this analysis on our three objects are presented in Table~\ref{lbolandeddington}.

\subsection{NGC 454E}
\label{ngc454lbol}

In the OM images, we identify the eastern source as being responsible for the AGN emission based on line ratio diagnostics using the \cite{1988A&A...191...29J} data and the other considerations discussed in \S\ref{ngc454_xray}.  In order to estimate the intrinsic extinction, we use the \cite{2001A&A...365...28M} dust-to-gas ratio (Fig. 1 of their paper).  Their paper shows hints that the ratio $\rm log(E(B-V)/N_{\rm H})$ reduces at low intrinsic X-ray luminosities, below $\sim10^{42} \rm erg/s$, with a sharp discontinuity leading to very high values of this ratio for even lower luminosities.   At the 2--10~keV luminosity of this source, there is a large uncertainty in the dust to gas ratio as given by \cite{2001A&A...365...28M}, and if we use $\rm log(E(B-V)/N_{\rm H}) = -23.6$ to $-23.0$ we find a range of values for the extinction $0.9<E(B-V)<3.68$.

This uncertainty does not, however, produce a correspondingly large uncertainty in the bolometric luminosity.  If we use $E(B-V)=0.9$, fit the combined X-ray and OM data with this value fixed into the dust extinction component, and integrate the de-reddened SED to determine the bolometric luminosity $L_{\rm bol}$, we find a value of $L_{\rm bol}=7.01\times 10^{43} \rm erg/s$.  Using the value of the intrinsic X-ray luminosity determined before, we find a bolometric correction to X-rays of 41, typical of that for Seyferts.  Employing the other extremal value of $E(B-V)=3.68$ only produces a 5\% change in $L_{\rm bol}$ and a very similar bolometric correction (38).  We therefore adopt $E(B-V)=0.9$ for all subsequent calculations.  The Eddington ratio estimate, using the \cite{2009ApJ...690.1322W} estimate of black hole mass (log $M_{\rm BH}$=6.23) is $\lambda_{\rm Edd}=0.32$.

Inspection of Hubble Space Telescope (\emph{HST}) \emph{Wide-Field Planetary Camera 2} (WFPC2, filter 814W at 8140$\rm \AA$) data for this source using the \textsc{ximage} software yields very little hint of any strong nuclear point source, as is expected for Seyfert 2 nuclei - the emission profile is entirely consistent with diffuse emission from the host galaxy.  So, we must consider the possibility that the OM emission, with much poorer resolution (larger point-spread-function, full-width half-maximum of $\sim 2''$) than the HST data (FWHM for the WFPC2 $\sim 0.1''$), represents a strong upper limit on the true AGN emission, and may give us a significant overestimate of the bolometric luminosity.   However, the presence of [Ne V] high-ionisation emission lines at 14.32 and 24.32$\rm \mu m$ in the nuclear mid-IR spectra of this object are a clear indication that AGN activity is present, and that there must be some emission from the central engine in the far-UV at $E>98 \rm eV$ to produce these lines \citep{2010ApJ...716.1151W}.  The severely reddened optical--UV SED shape is therefore highly likely to be indicative of dust.

For our second approach, we fit archival \emph{Spitzer} spectra with the host-galaxy and nuclear AGN templates from \cite{2004MNRAS.355..973S}.  We used Spitzer spectroscopy observations in the Short-High (SH) and Short-Low (SL) IRS order in the staring mode ($5.5 \lesssim \lambda <~ 19 \rm \mu m$). These spectroscopic apertures have the smallest angular resolution in the dispersion direction (3.6-4.7") and because of the slit positions in the sky, the SH and SL modules only include the nuclear contribution from NGC 454E. On the other hand, the extracted fluxes from the Long-High and Long-Low IRS order are contaminated with emission from the western companion galaxy, resulting in a continuum flux approximately twice the flux from the smaller slits alone.  We therefore restrict our fits to the short-wavelength SH and SL spectra.
 
We use the intrinsic X-ray luminosity and absorption $N_{\rm H}$ to select the appropriate IR SED templates to fit; for NGC 454E the column of $\sim 3 \times 10^{23} \thinspace \rm cm^{-2}$ implies that the nuclear template for log($N_{\rm H}$)=23--24 is most appropriate for fitting to the IR data.  However, the substantial variation in absorption in this source suggests that the Compton-thick template in \cite{2004MNRAS.355..973S} may also be appropriate.  However, we find that varying the choice of nuclear template used does not produce significantly different results for $L_{\rm bol}$.  The bolometric luminosity estimate from this method is $2.4 \times 10^{43} \rm erg \thinspace s^{-1}$.  The \cite{2012MNRAS.421.1803M} analysis of the \emph{Suzaku} data reveals a significantly different spectrum, and if we employ their best-fit model to determine the total 0.5--200 keV luminosity for the bolometric luminosity calculation, we find $L_{\rm bol} = 2.8 \times 10^{43} \rm erg \thinspace s^{-1}$.  Gathering together the different estimates of $L_{\rm bol}$ using the IR SED integration approach and using $\rm log(M_{\rm BH})=6.23$, this corresponds to an Eddington ratio of $0.11-0.13$, considering all possible X-ray spectral shapes (from XMM and Suzaku analyses) and column density estimates.  The bolometric correction to 2--10 keV using the XMM observation is $15$, whereas it reduces to $4$ if we use the results from the \emph{Suzaku} analysis.  Both the `direct' (optical--to--X-ray) estimation and the `reprocessed' measures of $L_{\rm bol}$ yield values less than the \cite{2009MNRAS.394L..89F} estimate, pushing this object further away from the `forbidden zone' of the $N_{\rm H}-\lambda_{\rm Edd}$ plot into the `long-lived absorption' region.   Notably, the IR-based estimate of $L_{\rm bol}$ is significantly less than the optical--to--X-ray SED integration estimate, indicating that there is much uncertainty in the dereddening correction and location of the turnover in the unobservable far-UV part of the spectrum.  For this source, the optical--to--Xray SED integration method may severly over-estimate the bolometric luminosity.

\subsection{2MASXJ03565655-4041453}

For this object, the intrinsic X-ray luminosity of $3.5 \times 10^{43} \rm erg/s$ implies $E(B-V)/N_{\rm H}$ is $\sim10^{-23} \rm mag \thinspace cm^{2}$ from the \cite{2001A&A...365...28M} plot, yielding an intrinsic extinction of $E(B-V)=0.71$.  Fitting the broad-band SED as before, using the \cite{2009ApJ...690.1322W} estimate of black hole mass to constrain the normalisation of the \textsc{diskpn} model, and hence, the position of the unobservable `turnover' in the big blue bump, we find a bolometric luminosity of $1.48\times 10^{45} \rm erg/s$.  For this object, we also have an estimate of the black hole mass using the V09/V10 refined method of dealing with the extended and nuclear K-band emission, and using this mass instead in the fit yields $L_{\rm bol}=1.83 \times 10^{45} \rm erg/s$, due to the different predicted turnover in the spectrum obtained using this mass estimate.  The Eddington ratios obtained using the two mass estimates are $0.03$ (W09 mass) and $0.06$ (V10 mass), with bolometric corrections of $42$ (W09 mass, $\rm log \thinspace M_{\rm BH} \thinspace =8.64$) or $52$ (V10 mass, $\rm log \thinspace M_{\rm BH} \thinspace =8.38$).  Combining results for the Eddington ratio and column density, we again find the object pushed away from the `forbidden region' into the `long-lived absorption' regime.   We cannot find any usable archival IR data for this source to estimate $L_{\rm bol}$ using the reprocessed emission.

\subsection{XSSJ05054-2348}

The intrinsic X-ray luminosity of $\sim 10^{43} \rm erg/s$ implies $E(B-V)/N_{\rm H}$ is $\sim10^{-23} \rm mag \thinspace cm^{2}$ from the \cite{2001A&A...365...28M} plot, yielding an intrinsic extinction of $E(B-V)=0.1$.  Fitting the broad-band SED as before, using the \cite{2009ApJ...690.1322W} estimate of black hole mass, we find a bolometric luminosity of $2.59\times 10^{44} \rm erg/s$.  We have another estimate of the black hole mass from V10 and using this mass instead in the fit yields $L_{\rm bol}=2.55 \times 10^{44} \rm erg/s$.  The Eddington ratios obtained using the two mass estimates are $0.06$ (W09 mass) and $0.04$ (V10 mass), with bolometric corrections of $8.6$ (W09 mass, $\rm log \thinspace M_{\rm BH} \thinspace = 7.53$) or $8.4$ (V10 mass, $\rm log \thinspace M_{\rm BH} \thinspace = 7.67$).

We also produce an IR-based estimate of the bolometric luminosity using archival \emph{Spitzer} and \emph{Herschel} data.  There is no confusion between the source of interest and any background sources in either dataset.  We used Spitzer spectroscopy observations in the Short-Low and Long-Low IRS order in the staring mode ($5.5 \lesssim \lambda \lesssim 36 \rm \mu m$).   We follow the same procedure to determine $L_{\rm bol}$ as for NGC 454E, described in \S\ref{ngc454lbol}, fitting host and nuclear components, and determining $L_{\rm bol}$ by adding the absorbed 0.5--200~keV luminosity to the nuclear IR luminosity from 8--1000$\rm \mu m$ with appropriate corrections for geometry and anisotropy (see \S\ref{sec:bollum}).  The bolometric luminosity from this method is $4.4 \times 10^{44} \rm erg \thinspace s^{-1}$.  This is a factor of 1.7 larger than the estimate from de-reddening the optical--UV SED and integrating, emphasising the uncertainty in the de-reddening of the optical--UV spectrum.  The corresponding 2--10~keV bolometric correction is 14.4 from this method, and the Eddington ratio estimates range between 0.07--0.12 (the range is from using the V10 mass or the W09 mass estimates).   Combining results for the Eddington ratio and column density, we again find the object pushed away from the `forbidden region' into the `long-lived absorption' regime, compared to the position of this object as presented in \cite{2009MNRAS.394L..89F}.

\begin{table*}
\begin{tabular}{llllllllllllllll}

\hline
AGN & $L_{\rm bol}^{(1)}$ & $L_{\rm bol}^{(2)}$ & $\kappa_{\rm 2-10 keV}^{(1)}$ & $\kappa_{\rm 2-10 keV}^{(2)}$ & $\lambda_{\rm Edd}^{(1)}$ & $\lambda_{\rm Edd}^{(2)}$ \\
 & (1) & (2) & (3) & (4) & (5) & (6) \\

\hline
NGC 454E & $7.01$ & $2.40$  & $41$ & $15$ & $0.32$ & $0.11$ \\
2MASX J03565655-4041453 & $148-183$ & -- & $42-52$ & -- & $0.03-0.06$ & -- \\
XSS J05054-2348 & $25.5-25.9$ & $44.0$ & $8.4-8.6$ & $14.4$ & $0.04-0.06$ & $0.07-0.12$ \\
\hline
\end{tabular}
\caption{SED fit results using XMM data in X-rays. All luminosities are in units of $10^{43} \rm erg \thinspace s^{-1}$. (1): bolometric luminosity from the optical--to--X-ray SED integration method (method 1). (2): bolometric luminosity from summing reprocessed nuclear IR emission and hard X-ray emission (method 2). (3): X-ray bolometric correction, $L_{\rm bol}/L_{\rm 2-10keV}$ using method 1 (4): X-ray bolometric correction from method 2. (5): Eddington ratio, $L_{\rm bol}/L_{\rm Edd}$, from method 1. (6): Eddington ratio from method (2). For objects where two estimates of the black hole mass (W09 and V10) are available, we quote ranges in derived quantities arising from using two different black hole masses.\label{lbolandeddington}}
\end{table*}

\section{Constraints on $L_{\rm bol}$ and $\lambda_{\rm Edd}$ from mid-IR high-ionisation lines}
\label{highionIRlines}

The work of \cite{2011ApJ...738....6M} provides a constraint on the broad-band SED shape and bolometric luminosity.  They generated photoionisation models to characterize the small dispersion found in correlations between high-ionisation mid-infrared emission-line ratios in AGN in order to predict the slope connecting the intrinsic emission at 1 Ry to the intrinsic emission at 1 keV, providing a very useful constraint on the unobservable extreme-UV (EUV) emission.  Other works have investigated the relationship between other high-ionisation IR lines and the unobservable UV emission in individual objects (e.g., \citealt{2009MNRAS.394L..16M} for Ark 564), but the \cite{2011ApJ...738....6M} study constitutes the first attempt to constrain the unobserved EUV continuum in the statistically significant BAT AGN sample, and provides a useful empirical tool for connecting the readily observable 1~keV emission to the unobservable EUV emission. This allows us to estimate whether our bolometric luminosity estimates in the previous sections are sensible, using simple physical constraints on the UV energy required to produce mid IR emission lines.   For NGC 454E, we are able to obtain the mid-IR line ratios from \cite{2010ApJ...716.1151W} and predict an energy index $\alpha=1.7$ for the power-law joining the 13.6~eV and 1~keV emission (for $F_{\rm \nu} \propto \nu^{-\alpha}$).  If we extrapolate from the intrinsic 1~keV emission using this spectral index, we see from Fig.~\ref{ngc454_sed} that the emission connects well with the predicted 13.6~eV emission from our accretion disc model fit using the de-reddened \textsc{diskpn} model.  Whilst the IR-based estimate of $L_{\rm bol}$ is significantly lower than the estimate from the UV \textsc{diskpn} fit, the high-ionisation line ratios would imply that the $L_{\rm bol}$ estimate from the UV model fit is consistent with the physical constraints from the high-ionisation mid-IR lines.

We do not have high-ionisation lines available in the mid-IR spectrum for XSS J05054-2348 or 2MASX J03565655-4041453, but \cite{2011ApJ...738....6M} identify that the range of 13.6~eV--1~keV energy indices is 1.5--2.0 for the BAT AGN sample.  For XSS J05054-2348, if we use the minimal index 1.5 which produces the minimal possible emission at 13.6~eV, we find that it far exceeds that predicted from the \textsc{diskpn} big blue bump fit (Fig.~\ref{xssj_sed}).  In this source, optical--to--X-ray SED fitting is likely to severely underestimate $L_{\rm bol}$, and the IR-based estimate is more likely correct.  For 2MASX J03565655-4041453, we show the range of possible values for the 13.6~eV emission using $1.5<\alpha<2.0$, and find that the de-reddened \textsc{diskpn} model fit lies well within these limits (Fig.~\ref{2masxj_sed}).

\section{Discussion}

\subsection{The Effective Eddington Limit Paradigm}
\label{effeddparadigm}

\begin{figure*}
    \includegraphics[width=10cm]{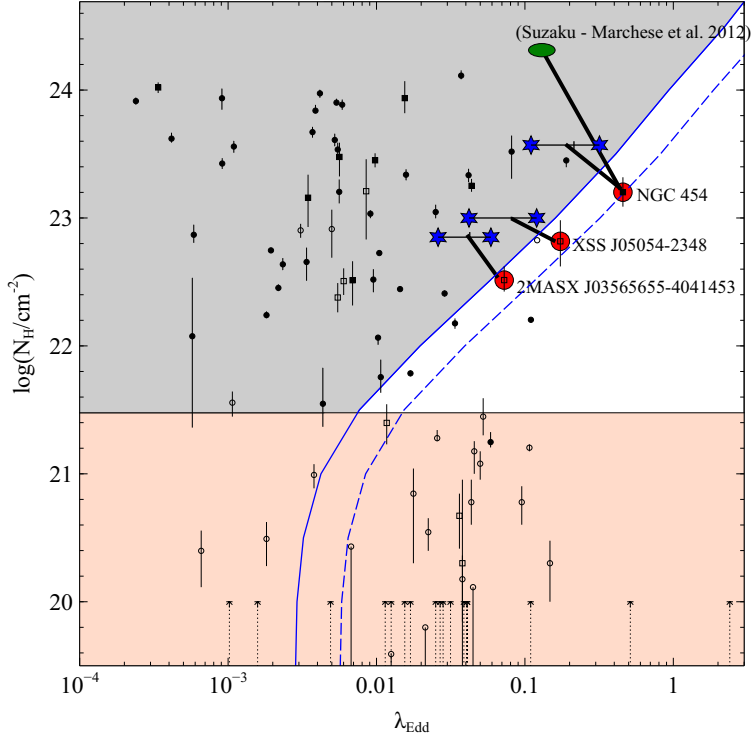}
    \caption{Changes in absorbing column and Eddington ratio for the three objects with new XMM data analysed in this paper.  Small circles represent objects with fits to archival ASCA or XMM-Newton data from Winter et al. (2009), and squares represent objects for which only XRT data were available at the time that W09 was published.  Black filled circles or squares depict objects for which the best fit was a `complex' spectrum, and empty circles or squares represent those with `simple' model fits.  Red circles show the positions in the $N_{\rm H}-\lambda_{\rm Edd}$ of the three objects with new data as given in \protect\cite{2009MNRAS.394L..89F} using XRT data only; lines connect these circles to the new positions of these objects as calculated from the new XMM data.  Blue stars show the extremes of the Eddington ratios that may be obtained using different black hole mass estimates, intrinsic reddening corrections or $L_{\rm bol}$ determination methods (i.e. optical--to--X-ray SED integration vs. using the IR reprocessed emission).  In all cases, the full range of Eddington ratios obtained, if combined with the new $N_{\rm H}$ estimates, place all three objects within the `long-lived absorption' region.\label{nhvsedd_withnew3}}

\end{figure*}

The refined estimates of column density and Eddington ratio using the high-quality XMM-Newton data suggest that all three objects under consideration here move away from the `forbidden zone' where short-lived absorption and outflows are expected.  These results are collated and summarised in Fig.~\ref{nhvsedd_withnew3}, along with the results for the rest of the 9-month BAT catalogue that were previously presented in \cite{2009MNRAS.394L..89F}.  We again draw the reader's attention to the other source (apart from the three studied in this paper) which was already known to lie well into the forbidden region, MCG-05-23-16, which is known to have a complex and variable warm absorber (e.g. \citealt{2007PASJ...59S.301R}) and a possible variable, high-ionisation, high-velocity outflow \citep{2007ApJ...670..978B}.   Other objects near the boundary are NGC 526A, which is found to have a warm absorber by \cite{2007MNRAS.382..194N}; NGC 3783, which has a known warm absorber from \cite{2001ApJ...554..216K}; and IC 4329A, where a prominent dust lane visible in optical images may be responsible for a large part of the $N_{\rm H}$ seen, and \cite{2006ApJ...646..783M} find an outflow.  Taken in conjunction with the discovery of \cite{2012ApJ...745..107W} (Fig.~13 of their paper) that warm absorber strength is highest near the Effective Eddington Limit, the evidence increasingly points to a scenario where the Effective Eddington Limit is of key importance in understanding the presence and nature of dusty gas absorption in AGN.  

Returning to the objects newly-observed with XMM studied in this paper, we note that in the case of NGC 454E, significant absorption variability between the XRT, XMM and Suzaku observations is observed, and the \cite{2012MNRAS.421.1803M} analysis of this XMM dataset comments on a distinct excess between 5--7~keV that could be indicative of an ionised absorber with significant column ($6 \times 10^{23} \rm cm^{-2}$), an ionisation parameter of log($\xi$)=3.6 with a turbulent velocity of 300 $\rm km \thinspace s^{-1}$, emerging from within the broad-line region.  However, we notice that the object has moved from the `forbidden region' (in the \emph{Swift}/XRT observation) to the `long-lived absorption' region with a Compton-thick column density (in the \emph{Suzaku} observation) to a less-absorbed state also in the `long-lived absorption' region, with an ionised absorber (in the \emph{XMM} observation). It is difficult to connect the properties of this ionised absorber directly with the Effective Eddington limit paradigm because of this unusual trajectory in parameter space.  

The class of AGN in which absorption variability is present (`changing look AGN') have been discussed extensively in the literature (e.g., \citealt{2010MNRAS.406L..20R,2009ApJ...705L...1R,2002ApJ...571..234R,2005A&A...442..185B}).  These AGN could potentially represent cases where radiation pressure acts to change the absorption on longer timescales, without necessarily manifesting as outflows.

\section{Summary and Conclusions}

Using high-quality XMM Newton spectra, we present new spectra for three objects from the 9-month Swift/BAT catalogue that were identified as being close to the Effective Eddington Limit for dusty gas.  The XMM spectra reveal that the best-fit model is a double-power law (partial covering absorption) for all three objects.  In each case, the scattered component in the second (low-energy) power-law for all three objects is $\sim$1 per cent of the primary power-law (emergent at high energies in the BAT spectra), consistent with simple electron scattering, not requiring any particularly sophisticated geometry for the absorption.

These spectra from XMM uncover hidden complexity in the spectrum for two of our objects, 2MASXJ03565655-4041453 and XSSJ05054-2348, for which the previously obtained XRT data were adequately fit by simple power-laws in the \cite{2009ApJ...690.1322W} analysis.  This highlights the usefulness of XMM in unearthing nuances in the spectra of objects, providing more information about the environment in these objects.

Taking into account uncertainties in the black hole mass estimates and estimates of intrinsic reddening, we have produced estimates of the bolometric luminosities and Eddington ratios using the new XMM data.  We employ the XMM OM data to generate broad-band SEDs for the three objects and attempt to recover their bolometric luminosities by applying a first-order reddening correction using a dust-to-gas conversion factor from the X-ray column density \citep{2001A&A...365...28M}.  We also use the re-processed infrared emission, as traced by \emph{Spitzer} and \emph{Herschel} observations where available, to obtain independent estimates of the bolometric luminosity and compare them with those obtained from optical--to--X-ray SED integration.  We use estimates of the black hole mass using the \cite{2009MNRAS.399.1553V} method if available; in the case of NGC 454E we only have one $M_{\rm BH}$ estimate (from W09).  The Eddington ratios determined all lie further into the `long-lived absorption' regime than found previously.  NGC 454E exhibits the highest variability in the absorption.  This AGN also exhibits the highest Eddington ratio of the three; the variable absorption places it in the `changing-look AGN' category \citep{2002ApJ...571..234R}, so it may be the case that the changing absorption is driven by radiation pressure from the central engine.  However, there is little evidence so far to suggest that changing-look AGN as a class are consistently located at the Effective Eddington limit.

This study reinforces the need to use high-quality spectra to determine the nature and amount of absorption in AGN, and re-emphasises the need for good mass estimates to determine Eddington ratios.  The recent comprehensive study 58-month BAT catalogue AGN in the Northern Galactic Cap ($b>50^{\circ}$, \citealt{2013ApJ...763..111V}) presents refined, systematically determined $N_{\rm H}$ estimates for the latest, deepest version of the BAT AGN catalogue.  If coupled with carefully determined black hole masses, this could provide a useful revision of the \cite{2009MNRAS.394L..89F} work on the earlier BAT catalogue, offering new insights on the interplay between radiation pressure and absorption in local AGN.

\section{Acknowledgements}

We thank the anonymous referee for comments and suggestions which significantly improved the paper.  RVV has been funded by NASA grant NNX09P89G for this project (associated with the \emph{XMM-Newton} proposal entitled `The Effective Eddington Limit for AGN'), and ACF thanks the Royal Society for support.  This research has made use of observations obtained with \emph{XMM-Newton}, an ESA science mission with instruments and contributions directly funded by ESA member states and the National Aeronautics and Space Administration (NASA).  This research has also made use of the NASA Extragalactic Database (NED) which is operated by the Jet Propulsion Laboratory, California Institute of Technology, under contract with NASA.

\bibliographystyle{mnras} 
\bibliography{effeddxmm}

\end{document}